\documentclass[12pt]{article}
\pdfoutput=1

\usepackage{amsmath,amsfonts,amssymb,epsfig}
\usepackage{slashed}
\usepackage[width=0.9\textwidth,margin={30pt,30pt},font=small,labelfont=bf]{caption}
\numberwithin{equation}{section}

\usepackage{hyperref}

\usepackage[style=phys,articletitle=true,biblabel=brackets,chaptertitle=false,pageranges=false,eprint=true]{biblatex}
\usepackage{bbold}
\usepackage{xcolor}
\hypersetup{
    colorlinks,
    linkcolor={red!0!black},
    citecolor={blue!50!black},
    urlcolor={blue!80!black}
}

\usepackage{graphicx}
\usepackage{cancel}

\usepackage{color}
\usepackage{bm}
\usepackage{multirow}
\usepackage{xspace}

\newcommand{\exclude}[1]{}
\DeclareMathOperator\arctanh{arctanh}

\newcommand{\ep}{\epsilon}

\newcommand{\im}{\mathrm{i}}

\newcommand{\ept}{\tilde{\ep}}

\begin{document}

\begin{flushright}
\end{flushright}
\begin{center}

\vspace{1cm}
{\large{\bf Open Superstring First Mass Level Effective Lagrangian: \\
Massive Spin-2 in an Electromagnetic Background}}

\vspace{1cm}

{ Karim Benakli$^{1,a}$ \let\thefootnote\relax\footnote{$^a$kbenakli@lpthe.jussieu.fr}, 
Cassiano A. Daniel$^{2,b}$ \footnote{$^b$c.daniel@unesp.br} 
and Wenqi Ke$^{1,c}$ \footnote{$^c$wke@lpthe.jussieu.fr}
}

\vspace{0.5 cm}

{
\emph{$^1$ Sorbonne Universit\'e, CNRS,\\ Laboratoire de Physique Th\'eorique et Hautes Energies, LPTHE, F-75005 Paris, France.
}}

{
\emph{$^2$  
ICTP South American Institute for Fundamental Research \\
Instituto de F\'{i}sica Te\'{o}rica, Universidade Estadual Paulista \\
Rua Dr. Bento Teobaldo Ferraz 271, 01140-070, S\~{a}o Paulo - SP, Brasil.
}}

\end{center}
\vspace{0.7cm}

\begin{abstract}
 \vskip 2mm 
  Minimal coupling leads to problems such as loss of causality if one wants to describe charged particles of spin greater than one propagating in a constant electromagnetic background.  Regge trajectories in string theory contain such states, so their study may allow us to investigate possible avenues to remedy the pathologies. We present here two explicit forms, related by field redefinitions, of the Lagrangian describing the bosonic states in the first massive level of open superstrings in four dimensions. The first one reduces, when the electromagnetic field is set to zero, to the Fierz-Pauli Lagrangian for the spin-2 mode. The second one is a more compact form which simplifies the derivation of a Fierz-Pauli system of equations of motion and constraints.
\end{abstract}

\newpage


\setcounter{footnote}{0}

\section{Introduction}
There are several motivations to study the propagation of spin-2 states in an electromagnetic field \cite{Dirac:1936tg,Fierz:1939ix}. This includes the possibility to (roughly) approximate hadronic resonances at sufficiently low energies, but also as a problem of mathematical interest. It is remarkable that the content of the first massive level of bosonic open strings can be reduced to a spin-2 state. The other string oscillation modes give rise to St\"uckelberg fields that can be gauged away. Therefore, the study of the propagation of the first massive level open bosonic string states  in an electromagnetic background allowed Argyres and Nappi to write the first, and only available, consistent causal Lagrangian for a charged spin-2 \cite{Argyres:1989cu}. Although this Lagrangian is causal only in $D=26$ dimensions, the corresponding form of the equations of motion and the constraints give an a priori consistent Fierz-Pauli system in arbitrary dimension, as pointed out by \cite{Porrati:2010hm,Porrati:2011uu}. The analysis of the equations of motion derived from the Virasoro algebra of open bosonic strings allows to generalize this system for any integer spin state bigger than one \cite{Porrati:2010hm}. String field theory has also been used to study the second massive level of bosonic open strings by \cite{Klishevich:1998sr} and leads to the action of a charged spin-3 state coupled to states with lower spins.

In this work, we are interested instead in the massive  states of the four-dimensional superstring. One question we address here is: what is the action describing the massive bosonic modes of the first excited level of the theory?  In the absence of electromagnetic background, we know that the theory is described by a Fierz-Pauli action for the spin-2 state, and similar free actions for the other lower spin fields. But because of the presence of many states, the theory is more complicated than that of the bosonic string. Indeed, a superspace action has been derived in \cite{Benakli:2021jxs}, and equations of motion have been obtained in the Lorenz gauge, generalizing the early work  \cite{Berkovits:1998ua,Berkovits:1997zd}. In particular, it shows couplings induced by the electromagnetic background between the fields of different spins present at this mass level. One thus expects from \cite{Benakli:2021jxs}, a complicated Lagrangian for physical states, and it has not yet been explicitly written. How can the action of the bosonic modes be written as a deformation trivially reducing to Fierz-Pauli in the absence of an electromagnetic background?  In fact, the Lagrangian is not unique as different choices of physical fields can be made. We will present two versions. The first one allows, by setting the electromagnetic field to zero, to directly find the Fierz-Pauli Lagrangian for the massive spin-2, as well as the free Lagrangian for other fields. The second one is a more compact form which allows to find the same spin-2 equation of motion and constraints.

Another question we address concerns the equations of motion and constraints. In \cite{Benakli:2021jxs}, these equations couple  fields of different spins. Is it possible to decouple these fields and obtain the equations corresponding to each field separately? We will see that this is possible at the cost of a complicated redefinition of the field describing the spin-2. In this work, we will not decouple the different fields at the Lagrangian level. In the superspace action with which we start, the lower spin fields play also a role in imposing constraints that allow to keep the right number of degrees of freedom in the presence of the electromagnetic background.

Many of the derivation steps of the Lagrangian, the equations of motion and constraints are long and tedious. They are not included in order to simplify the presentation and allow the reader to access the most important results. These calculations will be presented in detail elsewhere \cite{prep}.

\section{A deformation of the Fierz-Pauli Lagrangian}

An open string carries two charges $q_0$ and $q_\pi$ located at its ends. For the low energy effective field theory, only the total charge $Q = q_0 + q_\pi$ appears. We consider a charged string propagating in a constant electromagnetic background $F_{mn}$. The corresponding covariant derivative reads
\begin{equation}\begin{aligned}
  D_n =  (\partial -i Q F\cdot X)_n
    \end{aligned}\label{D_n}
\end{equation}
where $X$ is a space-time coordinate. The extended nature of strings causes the  electromagnetic field to enter all our formulas through the matrix \cite{Abouelsaood:1986gd}
\begin{equation}\begin{aligned}
  \epsilon =  \frac{\Lambda^2}{\pi} \left[\arctanh\left(\frac{\pi q_0 F}{\Lambda^2}\right) +\arctanh\left(\frac{\pi q_\pi F}{\Lambda^2}\right)\right]
    \end{aligned}\label{epsilon}
\end{equation}
where $\Lambda$ stands for the fundamental string scale and the covariant derivative \eqref{D_n} is traded for 
\begin{equation}\begin{aligned}
\mathfrak{D}_m =  -\im \,  \mathfrak{M}_{mn}\,  D^n ,  \qquad \left[\mathfrak{D}_m , \mathfrak{D}_n \right] = \im \epsilon_{mn}
    \end{aligned}\label{mathfrakD_n}
\end{equation}
with $\mathfrak{M}$  a  matrix that satisfies:
\begin{equation}\begin{aligned}
  \mathfrak{M} \cdot \mathfrak{M}^T= \frac{\epsilon}{Q F}
    \end{aligned}\label{Mmatrix}
\end{equation}
It is important to note that in all the manipulations we perform here, the only property of $\epsilon$ that is used is that it is an anti-symmetric tensor. The $F_{mn}$-dependence of $\epsilon$ is never explicitly invoked. Therefore, one could in particular take  $\epsilon_{mn} \rightarrow Q F_{mn}$ and $\mathfrak{D}_m \rightarrow D_m$. This means forgetting about the string origin of our Lagrangian and the resulting equations, and take the point-like particle limit.

At the first mass level of the superstring, the bosonic sector has 12 degrees of freedom (d.o.f.) of mass $M$, among which a spin-2 field  described by the rank-2 tensor $h_{mn}$. The symmetric, traceless and vanishing divergence conditions leave on-shell 5 degrees of freedom in $h_{mn}$. There is also a massive vector field $\mathcal{C}_m$ counting 3 d.o.f.. The 4 remaining d.o.f. consist in four scalar fields $\mathcal{M}_1, \mathcal{N}_1$, $A$ and $B$. We will write the Lagrangian as a sum of two pieces $\mathcal{L}=\mathcal{L}_1+\mathcal{L}_2$.

$\mathcal{L}_1$ contains the decoupled complex scalars $\mathcal{M}_1, \mathcal{N}_1$ and the massive vector $\mathcal{C}_m$:
\begin{equation}
    \begin{aligned}
        \mathcal{L}_1=&\bar{\mathcal{C}}^m\left( \mathfrak{D}^2-M^2\right)
    \mathcal{C}_m+\mathfrak{D}^m \bar{\mathcal{C}}_m\mathfrak{D}^n {\mathcal{C}}_n+2\im\ep_{mn}\bar{\mathcal{C}}^m\mathcal{C}^n
        \\&+{\mathcal{\bar{M}}}_1\left( \mathfrak{D}^2-M^2\right)\mathcal{M}_1+\bar{\mathcal{N}}_1\left( \mathfrak{D}^2-M^2\right)\mathcal{N}_1 \\
    \end{aligned}\label{Lagrangian-1.1}
\end{equation}
From this Lagrangian, we get the equations for the scalar fields:
\begin{equation}\begin{aligned}
&\left(\mathfrak{D}^2-M^2\right)\mathcal{M}_1=0,\quad    \left(\mathfrak{D}^2-M^2\right)\mathcal{N}_1=0
\end{aligned}\end{equation}
and for the massive vector boson:
\begin{equation}\begin{aligned}
&\left(\mathfrak{D}^2 -M^2\right)\mathcal{C}_m  -\mathfrak{D}_m\mathfrak{D}_n\mathcal{C}^n+2\im\epsilon_{mn}\mathcal{C}^n=0
,\quad \mathfrak{D}^m\mathcal{C}_m=0.
\end{aligned}\end{equation}

On the other hand, the spin-2 state ${h}_{mn}$ appears coupled to the complex scalars $A$ and $B$ in the second part of the Lagrangian. We define the re-scaled field:
\begin{equation}\begin{aligned}
     \mathcal{H}_{mn}\equiv \left( \eta_{mk}-\mathrm{i}\frac{2}{M^2}\epsilon_{mk}\right)h^{k}{}_n
    \end{aligned}\label{rescale-field-tensor}
\end{equation}
and denote  $\left(\ep\ep\right)\equiv\ep^{mn}\ep_{mn}$, $\left(\ep\ept\right)\equiv \ep^{mn}\ept_{mn}$, with the dual field strength $\ept_{mn}\equiv\varepsilon_{mnkl}\ep^{kl}/2$.
The Lagrangian of the remaining physical states $\{h_{mn}, A ,B\}$ reads:
\begin{equation}
\begin{aligned}  
\mathcal{L}_2=&\left[\bar{A}\mathfrak{D}_m-\frac{\im}{M^2}\tilde{\epsilon}_{mb}\bar{B}\mathfrak{D}^b+\frac{1}{2M^4}\left(\epsilon\tilde{\epsilon}\right)\bar{B}\mathfrak{D}_m  -\frac{1}{2}\varepsilon_{mabc}\mathcal{\bar{H}}^{bc}\mathfrak{D}^a-\frac{\im}{M^2}\tilde{\epsilon}_{ma}\mathcal{\bar{H}}^{ba}\mathfrak{D}_b+\frac{\im}{M^2}\tilde{\epsilon}_{mb}\bar{\mathcal{H}}\mathfrak{D}^b\right]\\&\quad   \times \left(\eta^{mn}-\frac{\im}{M^2}\epsilon^{mn}-\frac{2}{M^4}\tilde{\epsilon}^{mk}\tilde{\epsilon}_k{}^n\right)^{-1}
\\&
\times\left[\mathfrak{D}_n A +\frac{\im}{M^2}\tilde{\epsilon}_{nl}\mathfrak{D}^lB +\frac{1}{2M^4}\left(\epsilon\tilde{\epsilon}\right)\mathfrak{D}_nB  -\frac{1}{2}\varepsilon_{nlpq}\mathfrak{D}^l \mathcal{H}^{pq}+\frac{\im}{M^2}\tilde{\epsilon}_{nl}\mathfrak{D}_p\mathcal{H}^{pl}-\frac{\im}{M^2}\tilde{\epsilon}_{nl}\mathfrak{D}^l\mathcal{H}\right]
\\&-M^2\bar{A}A+\bar{B}\left(\mathfrak{D}^2-M^2\right)B-\frac{2}{M^4}\epsilon_{mn}\epsilon^{mk}\bar{B}\mathfrak{D}^n\mathfrak{D}_kB+ \frac{1}{2} \bar{\mathcal{H}}_{(mn)} \mathfrak{D} ^2h^{mn}+ \frac{1}{2}\mathfrak{D} ^{n} \bar{\mathcal{H}} _{mn} \mathfrak{D}_{k}h^{mk}\\&-M^2\bar{\mathcal{H}}^{(mn)}\mathcal{H}_{(mn)}+\frac{M^2}{2}\bar{\mathcal{H}}^{(mn)}{h}_{mn} -\frac{1}{2}\bar{\mathcal{H}} \left(\mathfrak{D}^2-M^2\right)h + \frac{1}{2}\mathfrak{D} ^{n} \bar{\mathcal{H}} _{nm} \mathfrak{D}_{k}\mathcal{H}^{km}\\&+\frac{1}{2}\left(\bar{\mathcal{H}}^{mn}\mathfrak{D}_m\mathfrak{D}_nh+\text{h.c.} \right)+\frac{1}{2M^2}\left[2\mathrm{i}\left(\mathfrak{D}^n\bar{\mathcal{H}}_{nm}\epsilon^{mk}\mathfrak{D}_kB \right)-\left(\ep\ep\right)\bar{\mathcal{H}}B+\text{h.c.} \right]
\\& +\frac{M^2}{2}\left(\bar{\mathcal{H}}^{[mn]}+\frac{1}{M^2}\mathrm{i}\epsilon^{mn}\bar{B}\right)\left(\mathcal{H}_{[mn]}-\frac{1}{M^2}\mathrm{i}\epsilon_{mn}B\right)
\end{aligned}\label{Lagrangian-1.2}
\end{equation}

Despite the complicated couplings in \eqref{Lagrangian-1.2}, the different fields can be decoupled in the equations of motion. Compact expressions of the latter are obtained after some algebraic manipulations, which do not involve redefining the fields, of the Euler-Lagrange equations. From (\ref{Lagrangian-1.2}), we can thus derive the following equations for the spin-0 fields:
\begin{equation}\begin{aligned}
   \left(\mathfrak{D}^2-M^2\right)A=0,\quad    \left(\mathfrak{D}^2-M^2\right)B=0,
\end{aligned}\end{equation}
and for massive spin-2 field:
\begin{equation}\left(\mathfrak{D}^2-M^2 \right){h}_{mn}+2\im\left(\ep_{m}{}^k{h}_{nk}+\ep_{n}{}^k{h}_{mk}\right)=0 , \qquad h=0
\end{equation}
Unlike the equations of motion, the divergence constraint of the symmetric rank-2 tensor $h_{mn}$ is coupled to the scalar fields.
The divergence constraint reads:
\begin{equation}
    \begin{aligned}&\left[M^4-\frac{1}{2}\left(\ep\ep\right)\right]\mathfrak{D}^nh_{mn}-{\im}M^{2}\ep_{mk}\mathfrak{D}_nh^{nk}-{\im}M^{2}\ep^{nk}\mathfrak{D}_nh_{mk}+{2}\ep_{mk}\ep_{nl}\mathfrak{D}^lh^{nk} -\ep_{mk}\ep^{kn}\mathfrak{D}^lh_{nl}\\&\quad-\ep^{kn}\ep_{kl}\mathfrak{D}^lh_{mn}+\ep^{kn}\ep_k{}^l\mathfrak{D}_mh_{nl}+\frac{1}{2}\left(\ep \ep\right)\mathfrak{D}_mB +{\im}M^{2}\ep_{mn}\mathfrak{D}^nB+{2}\ep_{mk}\ep^{kn}\mathfrak{D}_nB-\im M^2 \tilde{\epsilon}_{mn}\mathfrak{D}^nA\\&=0
    \end{aligned}\label{Vprime}
\end{equation}
which for $\ep=0$ becomes $\partial^nh_{mn}=0$. Eq.~\eqref{Vprime} includes many $\ep$-dependent terms and couplings to the scalars, however, there is an on-shell redefinition of the spin-2:
\begin{equation}
\begin{aligned} \mathfrak{h}_{mn}\equiv&\frac{2}{3}h_{mn}-\frac{\im}{M^2}\ep_{m}{}^kh_{kn}-\frac{1}{3M^2}\mathfrak{D}_m\mathfrak{D}_kh^{k}{}_n
\\&
+\frac{\im}{M^4}\left(\ep_{mk}\mathfrak{D}^l\mathfrak{D}^kh_{nl}-\ep_{mk}\mathfrak{D}_l\mathfrak{D}_nh^{kl}-\frac{1}{2}\eta_{mn}\ep^{kl}\mathfrak{D}^p\mathfrak{D}_lh_{kp}\right)\\&+\frac{1}{2M^4-4\ep\ep}\left[\frac{4}{3M^2}\left(\ep\ep\right)\mathfrak{D}_m\mathfrak{D}_nB+\left(1-\frac{4}{M^4}\ep\ep\right)\ep_m{}^k\ep_{kn}B+{2\im}\ep_{mk}\mathfrak{D}^k\mathfrak{D}_nB\right.\\&\quad \qquad \qquad \qquad \left.-\left(\frac{1}{3}+\frac{1}{M^4}\ep\ep\right)\left(\ep\ep\right)\eta_{mn}B\right]\\&+\frac{1}{2M^4+4\ep\ep}\left[-{2\im}\ept_{mk}\mathfrak{D}^k\mathfrak{D}_nA+\frac{5}{4}\left(\ep\ept\right)\eta_{mn}A -\frac{2}{M^2}\left(\ep\ept\right)\mathfrak{D}_m\mathfrak{D}_nA+\frac{8}{M^2}\ept_{mk}\ep_{ln}\mathfrak{D}^k\mathfrak{D}^l A\right ]\\&+\left(m\leftrightarrow n\right)
    \end{aligned}\label{sp2ABdef}
\end{equation}
which decouples the divergence constraint to give the deformed Fierz-Pauli system:
\begin{equation}
    \begin{aligned}
        &\left(\mathfrak{D}^2-M^2 \right)\mathfrak{h}_{mn}+2\im\left(\ep_{m}{}^k\mathfrak{h}_{nk}+\ep_{n}{}^k\mathfrak{h}_{mk}\right)=0 \\&\mathfrak{D}^n\mathfrak{h}_{mn}=0,\quad \mathfrak{h}=0 
\end{aligned}
\end{equation}

Therefore, in $\mathcal{L}_1$ and $\mathcal{L}_2$ one starts with 18 complex degrees of freedom. Taking into account the constraints on the spin-2 and spin-1 fields, this leaves us with 12 of them on shell. Introducing the redefinition \eqref{sp2ABdef}, all equations of motion and constraints  are   decoupled. When $\epsilon=0$, \textit{i.e.}~without background,  $\mathcal{L}_2$  is decoupled, and we recover  two scalars $A$, $B$ as well as a spin-2 $h_{mn}$ described by the usual Fierz-Pauli Lagrangian.

\section{A compact Lagrangian}
With $\mathcal{L}_1$ unchanged, we present here an equivalent, but more compact version of $\mathcal{L}_2$, which allows a simpler derivation of the equations of motion and constraints. The following Lagrangian involves two vectors $\{a_m,c_m\}$ instead of the scalars $\{A, B\}$. There is indeed a duality relation between the two sets of fields \cite{Curtright:1980yj}, because only the longitudinal components of the vectors are physical degrees of freedom, the other components being projected out by the constraints. The Lagrangian reads:
\begin{equation}
\begin{aligned}
\mathcal{L}_2=&\bar{a}^m\left(M^2\eta_{mn}-\mathrm{i}\epsilon_{mn}\right) a^n+\mathfrak{D}^m\bar{a}_m\mathfrak{D}^n a_n-M^2\bar{c}^m c_m-\frac{2}{5}\mathfrak{D}^m \bar{c}_m\mathfrak{D}^n c_n\\&+\frac{1}{\sqrt{2}}\left[M\bar{c}^m\left(-\frac{2}{5}\mathfrak{D}_m\mathcal{H}+\mathfrak{D}^n\mathcal{H}_{nm}\right)+\bar{\tilde{F}}^{mn}(a)\left(F_{mn}(c)-\frac{M}{\sqrt{2}}\mathcal{H}_{[mn]}\right)+\text{h.c.}\right]\\&+ \frac{1}{2} \bar{\mathcal{H}}_{mn} \mathfrak{D} ^2h^{mn}+ \frac{1}{2}\mathfrak{D} ^{n} \bar{\mathcal{H}} _{mn} \mathfrak{D}_{k}h^{mk}-\frac{M^2}{2}\bar{\mathcal{H}}^{(mn)}\mathcal{H}_{(mn)}+\frac{M^2}{20} \bar{\mathcal{H}}\mathcal{H}+\mathrm{i}\epsilon^{nk}\bar{\mathcal{H}}_{mn}h_{k}{}^m\end{aligned}\label{compact-boson}
\end{equation}
with the (dual) field strengths given by
\begin{equation}
    F_{mn}(a)\equiv \mathfrak{D}_m a_n- \mathfrak{D}_n a_m,\quad \tilde{{F}}_{mn}(a)\equiv \frac{1}{2}\varepsilon_{mnpq}F^{pq}(a)
\end{equation}
and similarly for $F_{mn}(c),\tilde{F}_{mn}(c)$. 

We will derive here the equations of motion and the constraints on the fields. To analyze the Lagrangian, we first decompose $h_{mn}$ into its trace $h$ and a traceless part $v_{mn}$: $h_{mn}=\frac{1}{4}\eta_{mn}h+v_{mn}$. Combining the equations of motion of $h$ and $v_{mn}$, we obtain:
\begin{equation}
5\mathfrak{D}^2h=3M^2h-8\sqrt{2}M\mathfrak{D}^m c_m\label{d2hrelation}
\end{equation}

The  Euler-Lagrange equations of $a_m$, $c_m$  read
\begin{equation}
    \begin{aligned}&
M^2a_m-\mathrm{i}\epsilon_{mn}a^n-\mathfrak{D}_m\mathfrak{D}_n a^n+\sqrt{2}\mathrm{i}\tilde{\epsilon}_{mn}c^n-\frac{\mathrm{i}}{M} \epsilon^{nk}\varepsilon_{mkpq}\mathfrak{D}^q h_{n}{}^p=0\\   &  -M^2c_m +\frac{2}{5}\mathfrak{D}_m\mathfrak{D}_nc^n+\sqrt{2}\mathrm{i}\tilde{\epsilon}_{mn}a^n-\frac{\sqrt{2}}{5}M\mathfrak{D}_mh+ \frac{1}{\sqrt{2}M}\left(M^2\eta^{nk}-2\mathrm{i}\epsilon^{nk}\right)\mathfrak{D}_nh_{mk}=0
    \end{aligned}
\label{eomc}\end{equation}

It is useful to compute their divergences
\begin{equation}
\begin{aligned}
0=&M^2\mathfrak{D}^ma_m-\mathrm{i}\epsilon_{mn}\mathfrak{D}^ma^n-\mathfrak{D}^2\mathfrak{D}_ma^m+\sqrt{2}\mathrm{i}\tilde{\epsilon}_{mn}\mathfrak{D}^mc^n
-\frac{1}{4M}\left(\ep\ept\right) h
 \\  0=&  -M^2\mathfrak{D}^mc_m +\frac{2}{5}\mathfrak{D}^2\mathfrak{D}_mc^m+\sqrt{2}\im\tilde{\epsilon}_{mn}\mathfrak{D}^ma^n-\frac{\sqrt{2}}{5}M\mathfrak{D}^2h\\&+\frac{1}{\sqrt{2}M}\left(M^2\eta^{nk}-2\mathrm{i}\epsilon^{nk}\right)\mathfrak{D}^m\mathfrak{D}_nh_{mk}
\end{aligned} \label{Div-am-cm}   
\end{equation}


Next, we consider the Euler-Lagrange equation  of $\mathcal{H}_{mn}$:
\begin{equation}
    \begin{aligned}
 \mathcal{R}_{mn}\equiv    & -\frac{1}{M}\varepsilon_{mnkl}\mathfrak{D}^k a^l +\frac{2\sqrt{2}}{5M}\eta_{mn}\mathfrak{D}^k c_k -\frac{\sqrt{2}}{M}\mathfrak{D}_m c_n
 +\frac{1}{10}\eta_{mn}h+\frac{1}{M^2}\mathfrak{D}^2 h_{mn}\\&-\frac{1}{M^2}\mathfrak{D}_n\mathfrak{D}^k h_{mk}-h_{mn}-\frac{3}{M^2}\mathrm{i}\epsilon_{kn}h^k{}_m+\frac{1}{M^2}\mathrm{i}\epsilon_m{}^k h_{kn}\\&=0
    \end{aligned}
\end{equation}
whose trace gives\begin{equation}
    \begin{aligned}
  \mathcal{R}^m{}_{m}=    \frac{3\sqrt{2}}{5M}\mathfrak{D}^m c_m -\frac{3}{5}h +\frac{1}{M^2}\mathfrak{D}^2 h -\frac{1}{M^2}\mathfrak{D}^m\mathfrak{D}^n h_{mn}=0
    \end{aligned}
\end{equation}
and using the relation \eqref{d2hrelation}, it implies 
\begin{equation}
    \mathfrak{D}^m c_m +\frac{1}{\sqrt{2}M}\mathfrak{D}^m \mathfrak{D}^n h_{mn}=0\label{doublediv}
\end{equation}
Having \eqref{d2hrelation}, \eqref{Div-am-cm} and \eqref{doublediv} at hand, the trace constraint of $h_{mn}$ can be now inferred from  the sum $\left(\mathrm{i}\epsilon^{mn}\mathcal{R}_{mn}+   \mathfrak{D}^m\mathfrak{D}^n \mathcal{R}_{mn} \right)$:
\begin{equation}
    Mh+4\sqrt{2}\mathfrak{D}^mc_m=0\label{tracediv}
\end{equation}

The next step involves calculating the divergence constraint of $h_{mn}$. To this end, let us contract  $\mathcal{R}_{mn}$ with $\mathfrak{D}^n$, and simplify the result using the equation of motion~of $c_m$:
\begin{equation}
    \begin{aligned}
    0=  \mathfrak{D}^n \mathcal{R}_{mn}=&-\frac{M}{\sqrt{2}}c_m -\frac{2\sqrt{2}}{5M} \mathfrak{D}_m\mathfrak{D}_nc^n +\mathrm{i}\frac{\sqrt{2}}{M}\epsilon_{mn}c^n\\&-\frac{1}{10}\mathfrak{D}_m h -\frac{1}{2}\mathfrak{D}^n h_{mn}+\frac{1}{M^2}\mathrm{i}\epsilon_{mk} \mathfrak{D}_nh^{kn}   \end{aligned}
\end{equation}
Moreover, thanks to  the equation \eqref{tracediv}, the term $\mathfrak{D}_m\mathfrak{D}_nc^n$ above can be replaced by $-\frac{M}{4\sqrt{2}}\mathfrak{D}_m h$, which results in
\begin{equation}
    \begin{aligned}
    \mathfrak{D}^n \mathcal{R}_{mn}=&-\frac{M}{\sqrt{2}}\left( \eta_{mn}-\frac{2\im}{M^2}\epsilon_{mn}\right)c^n -\frac{1}{2}\left( \eta_{mk}-\frac{2\im}{M^2}\epsilon_{mk}\right)\mathfrak{D}_nh^ {kn}
    \\& =0
    \end{aligned}
\end{equation}
Applying the inverse matrix $\left( \eta_{mn}-2\mathrm{i}\epsilon_{mn}/M^2\right)^{-1}$, this equation implies the divergence constraint\begin{equation}
    \begin{aligned}
  c_m+\frac{1}{\sqrt{2}M}\mathfrak{D}^n h_{mn}=0 
   \end{aligned}
\end{equation}

Coming back to the original equation of motion~ $\mathcal{R}_{mn}$, its symmetric and antisymmetric parts can now be rewritten as, respectively:
\begin{equation}
    \begin{aligned}
      2M^2\mathcal{R}_{(mn)}=&\left( \mathfrak{D}^2-M^2\right)h_{mn}-2\mathrm{i}\left( \epsilon_{km}h^k{}_n+\epsilon_{kn}h^k{}_m\right)=0
\\     2M^2\mathcal{R}_{[mn]}=&-{2}{M}\varepsilon_{mnkl}\mathfrak{D}^k a^l -{2\sqrt{2}}{M}\left( \mathfrak{D}_mc_n -\mathfrak{D}_nc_m\right)+{2\im}\left(  \epsilon_{km}h^k{}_n-\epsilon_{kn}h^k{}_m\right)=0
\end{aligned}\label{Sym-Antisym}
\end{equation}

The first line in \eqref{Sym-Antisym} is precisely the equation of motion~of the Fierz-Pauli system and it is the four-dimensional version of the same form  obtained in \cite{Argyres:1989cu} for $D=26$ dimensions. The second line establishes a duality relation between the field strength of $a_m$, and the sum of the field strength of $c_m$ and $\mathcal{H}_{[mn]}=\frac{\mathrm{i}}{M^2}\left(  \epsilon_{km}h^k{}_n-\epsilon_{kn}h^k{}_m\right)$.

To summarize, for the symmetric tensor $h_{mn}$, we have derived the following equations of motion~and constraints
\begin{equation}
    \begin{aligned}
        &\left( \mathfrak{D}^2-M^2\right)h_{mn}-2\mathrm{i}\left( \epsilon_{km}h^k{}_n+\epsilon_{kn}h^k{}_m\right)=0\\&\mathfrak{D}^nh_{mn}+\sqrt{2}Mc_m=0,\quad Mh=-4\sqrt{2}\mathfrak{D}^mc_m
    \end{aligned}
\end{equation}
whereas the equations of motion of the vectors $\{a_m, c_m\}$, given by \eqref{eomc}, are still coupled.  By considering the independent constraints obtained so far, the d.o.f.~counting will reveal 12 complex d.o.f.'s on shell. To make this  manifest, we start by decoupling the equations of motion of $\{a_m,c_m\}$. With the on-shell redefinitions
\begin{equation}
    \begin{aligned}&
        a_m^\prime\equiv a_m -\frac{\im}{M^2}\epsilon_{mn}a^n -\frac{\im }{M^3}\ept^{nk}\mathfrak{D}_kh_{mn}+\frac{\im}{M^3}\ept_{mn}\mathfrak{D}^n h +\frac{2\sqrt{2}}{M^2}\im \ept_{mn}c^n \\&c_m^\prime\equiv c_m-\frac{\sqrt{2}\im}{2M^2}\ept_{mn}a^n +\frac{\im}{\sqrt{2}M^3}\ep^{nk}\mathfrak{D}_nh_{mk}
    \end{aligned}\label{newamcm}
\end{equation}
their equations of motion become
\begin{equation}
    \mathfrak{D}_m\mathfrak{D}_na^{\prime n}=M^2a^\prime_m,\quad     \mathfrak{D}_m\mathfrak{D}_nc^{\prime n}=M^2c^\prime_m
\end{equation}
indicating that $a_m^\prime$, $c_m^\prime$ are the gradient of a scalar and as such have only one (longitudinal) d.o.f.~on shell. More precisely, they are dual to the massive scalars $\{A,B\}$ in the previous section, in the same manner as described in \cite{Curtright:1980yj}.

To decouple the trace constraint of $h_{mn}$, a natural guess would be:  \begin{equation}\begin{aligned}\mathcal{H}_{mn}^{\prime}&\equiv \mathcal{H}_{(mn)}+\frac{\sqrt{2}}{M}\eta_{mn}\mathfrak{D}^k c_k
\end{aligned}
\end{equation}
where $\mathcal{H}_{(mn)}$ is the symmetric part of the rescaled $h_{mn}$, defined in \eqref{rescale-field-tensor}. This new spin-2 satisfies
\begin{equation}
\begin{aligned}
&\left( \mathfrak{D}^2-M^2\right)\mathcal{H}^{\prime}_{mn}+2\mathrm{i}\left[\left(\epsilon\cdot \mathcal{H}^{ \prime}\right)_{mn}-\left( \mathcal{H}^{\prime} \cdot \epsilon\right)_{mn} \right]=0,
\\& \mathfrak{D}^n\mathcal{H}^{\prime}_{mn}=-\frac{\mathrm{i}}{M}\tilde{\epsilon}_{mn}a^n +\mathrm{i}\frac{\sqrt{2}}{M}\epsilon_{mn}c^n
,\quad  \mathcal{H}^{\prime}=0
\end{aligned}
\end{equation}
In the free case, $\ep=0$, the above equations reduce to the Fierz-Pauli system. The divergence constraint of $\mathcal{H}^\prime_{mn}$ is coupled to $\{a_m,c_m\}$ with the field strength $\ep_{mn}$, which resembles the result in \cite{Benakli:2021jxs}, Eq (6.72), but $a_m$ in this reference is a massive vector boson having 3 d.o.f.'s.

We may also attempt to use the redefinition in \cite{Argyres:1989cu}, where the only physical field is the charged spin-2. It was convenient to introduce the double rescaling:\begin{equation}
    \mathcal{H}^{\prime\prime}_{mn}= \left(\eta_{mk}-\mathrm{i}\frac{2}{M^2}\epsilon_{mk} \right)\left(\eta_{nl}-\mathrm{i}\frac{2}{M^2}\epsilon_{nl} \right)h^{kl}\label{Hmn1}
\end{equation}This ansatz can be modified to have a vanishing trace in our case:\begin{equation}
    \begin{aligned}
    \mathcal{H}^{\prime\prime}_{mn}=& \left(\eta_{mk}-\mathrm{i}\frac{2}{M^2}\epsilon_{mk} \right)\left(\eta_{nl}-\im\frac{2}{M^2}\epsilon_{nl} \right)h^{kl}
     \\& +\frac{\sqrt{2}}{2M}\eta_{mn}\left(2\mathfrak{D}^k c_k
    +\mathrm{i}\frac{\sqrt{2}}{M^2}\tilde{\epsilon}^{kl}\mathfrak{D}_k a_l +\frac{2}{M^2}\mathrm{i}\epsilon^{kl}\mathfrak{D}_kc_l\right),\\ 
    \mathcal{H}^{\prime\prime}=&0
   \end{aligned}  \label{Hmn2}
  \end{equation}
  But again, the divergence constraint is non-vanishing. And to absorb the latter, other terms of order $\mathcal{O}(\ep^2)$ and higher derivatives shall be included in the new spin-2 definition. Beginning with a general ansatz that has such higher order terms, and imposing the divergence as well as the trace constraint, we find a redefinition that satisfies these requirements:
  \begin{equation}
    \begin{aligned}
   \mathfrak{h}_{mn}\equiv&\frac{2}{3}h_{mn}-\frac{1}{6}\eta_{mn}h-\frac{\im}{M^2}\ep_{m}{}^kh_{kn}+\frac{\sqrt{2}}{3M}\mathfrak{D}_mc_n-\frac{1}{M^4}\left(\ep_{mk}\ep^{lk}h_{nl}+\ep_{mk}\ep_{nl}h^{kl}-\frac{1}{2}\eta_{mn}\ep^{kl}\ep^p{}_lh_{kp}\right)\\&-\frac{\im\sqrt{2}}{M^3}\left(\ep_{mk}\mathfrak{D}^kc_n-\ep_{mk}\mathfrak{D}_nc^k+\frac{1}{2}\eta_{mn}\ep^{kl}\mathfrak{D}_kc_l\right)\\&-\frac{1}{2M^4+4\ep\ep}\left[-\frac{2\im}{M}\ept_{mk}\mathfrak{D}^k\mathfrak{D}_n\mathfrak{D}_la^l+\frac{5}{4M}\left(\ep\ept\right)\eta_{mn}\mathfrak{D}^ka_k -\frac{2}{M^3}\left(\ep\ept\right)\mathfrak{D}_m\mathfrak{D}_n\mathfrak{D}_ka^k\right.\\&\qquad\qquad\qquad   \left.+\frac{8}{M^3}\ept_{mk}\ep_{ln}\mathfrak{D}^k\mathfrak{D}^l \mathfrak{D}^pa_p\right ]\\&+\frac{1}{2M^4-4\ep\ep}\left[\frac{M^2}{6}\mathfrak{D}_m \mathfrak{D}_n h-\frac{1}{4}\ep_{mk}\ep^k{}_nh+\frac{\im}{2}\ep_{mk}\mathfrak{D}^k\mathfrak{D}_nh-\frac{5\ep\ep}{24}\eta_{mn}h\right]+\left(m \leftrightarrow n\right)
    \end{aligned}\label{sp2ABdef2}
\end{equation}
One can check that the above definition yields a decoupled Fierz-Pauli system. To recapitulate, with the redefined fields \eqref{newamcm} and \eqref{sp2ABdef2}, the Lagrangian \eqref{compact-boson} gives on-shell the following equations of motion and constraints:
\begin{equation}
\begin{aligned}
&\left( \mathfrak{D}^2-M^2\right)\mathfrak{h}_{mn}=2\mathrm{i}\left( \epsilon_{km}\mathfrak{h}^k{}_n+\epsilon_{kn}\mathfrak{h}^k{}_m\right)\\& \mathfrak{D}^n\mathfrak{h}_{mn}=0,\quad \mathfrak{h}=0\\
&\mathfrak{D}_m\mathfrak{D}_na^{\prime n}=M^2a^\prime_m,\quad     \mathfrak{D}_m\mathfrak{D}_nc^{\prime n}=M^2c^\prime_m
\end{aligned}
\end{equation}
Finally, we can proceed to an immediate count of the degrees of freedom. We find the presence of 12 on-shell bosonic degrees of freedom described by $\mathcal{L}_1+\mathcal{L}_2$. Although, contrary to \eqref{Lagrangian-1.2}, $\mathcal{L}_2$ given by \eqref{compact-boson} involves extra non-propagating fields, the transverse components of $\{a_m^\prime,c_m^\prime\}$, it has the advantage of a more compact form and allows a simpler derivation of equations of motion and constraints.

\section{Conclusions}
We have presented here for the first time the explicit expression of the two-derivative Lagrangian describing the states of the first excited level of the open superstring in the presence of a constant electromagnetic background. In fact, we have given two different forms, each with its own advantages.

One result of our work that we want to comment on is that the equations of motion and constraints obtained are those expected. In particular, for the massive spin-2, we find at the end a Fierz-Pauli system of the form described by Argyres-Nappi, which is what would be expected if it were a dimensional reduction. We want to emphasize, however, that the intermediate calculations are very different. First of all, at the computational level, let us note that unlike the bosonic case, for superstrings the mass does not depend on the electromagnetic background. Then, more importantly, the first mass level contains other physical fields than the massive spin-2 and the equations of motion directly derived from the Lagrangian couple the different fields. But we have explicitly shown how we can decouple them at the level of these equations, which is not trivial. This required a redefinition of the fields which can be implemented at the level of the Lagrangian to obtain directly the decoupled equations. We perform this procedure in another work devoted to fermionic partners, but here the expressions become long and complicated, which is not desirable. Let us finally note that for all Lagrangians, presented here and in the previous literature, the Euler-Lagrange equations do not give directly the equations in their final compact forms but algebraic manipulations are necessary to reduce them to these forms. This also highlights the difficulty of guessing a four-dimensional Lagrangian for the charged massive spin-2 from the equations of motion.

\section*{Acknowledgements}

We are grateful to Nathan Berkovits for useful discussions.
CAD acknowledges FAPESP grant numbers 2020/10183-9 and 2022/14599-0 for financial
support. The work of WK is
supported by the Contrat Doctoral Sp\'ecifique Normalien (CDSN) of Ecole Normale Sup\'erieure –
PSL.

\printbibliography

\end{document}